\documentclass[a4paper]{jpconf}
\usepackage{graphicx}
\usepackage{graphicx}
\begin{document}
\newcommand{\bqq}{\begin{eqnarray}}
\newcommand{\eqq}{\end{eqnarray}}
\newcommand{\PbPb}     {Pb--Pb}
\newcommand{\cms}{\sqrt{s}}
\newcommand{\cmsNN}{\sqrt{s_{\rm NN}}}
\newcommand{\cmsofT}[1]{$\sqrt{s} = \unit[#1]{TeV}$}
\newcommand{\cmsofG}[1]{$\sqrt{s} = \unit[#1]{GeV}$}
\newcommand{\expval}[1]{\langle #1 \rangle}
\newcommand{\etain}[1]{$|\eta|$~$<$~$#1$}
\newcommand{\vtxin}[1]{$|\textrm{vt-x-}z| < \unit[#1]{cm}$}
\newcommand{\qbar}{\bar{q}}
\newcommand{\n}{\expval{n}}
\newcommand{\N}{\expval{N}}
\newcommand{\pbar}{\bar{\mbox{p}}}
\newcommand{\jt}{\ensuremath{j_{\rm{t}}}}
\newcommand{\pT}{\ensuremath{p_{\rm T}}}
\newcommand{\kt}{\ensuremath{k_{T}}}
\newcommand{\et}{\ensuremath{E_{\rm T}}}
\newcommand{\pta}{\ensuremath{p_{\rm t, assoc}}}
\newcommand{\ptt}{\ensuremath{p_{\rm t, trig}}}
\newcommand{\ptmin}{p_{\rm{t}, \textrm{min}}}
\newcommand{\ptcutoff}{p_{\textrm{t, cut-off}}}
\newcommand{\nch}{N_{\rm{ch}}}
\newcommand{\Dphi}{\Delta\varphi}
\newcommand{\Deta}{\Delta\eta}
\newcommand{\Ntrig}{N_{\rm trig}}
\newcommand{\Nassoc}{N_{\rm assoc}}
\newcommand{\icp}{I_{\rm CP}}
\newcommand{\iaa}{I_{\rm AA}}
\newcommand{\dd} {\mbox{${\rm d}$}}
\newcommand{\mev}     {\mbox{${\rm MeV}$}}
\newcommand{\gev}     {\mbox{${\rm GeV}$}}
\newcommand{\tev}     {\mbox{${\rm TeV}$}}
\newcommand{\gmom}    {\mbox{${\rm GeV}/c$}}
\newcommand{\mmom}    {\mbox{${\rm MeV}/c$}}
\newcommand{\df}      {\mbox{${\rm d}$}}
\newcommand{\raa}     {\mbox{$R_{\rm AA}$}}
\newcommand{\rcp}     {\mbox{$R_{\rm CP}$}}
\newcommand{\jpsi}    {\mbox{$J/\psi$}}
\newcommand{\lum}     {\mbox{${\rm cm}^{-2} {\rm s}^{-1}$}}
\newcommand{\RAA}     {\mbox{$R_{\rm AA}$}}
\newcommand{\RpA}     {\mbox{$R_{\rm pPb}$}}
\newcommand{\QpA}     {\mbox{$Q_{\rm pPb}$}}
\newcommand{\meanpt}    {\mbox{$\langle \pT \rangle$}}
\newcommand{\ncoll}    {\mbox{$N_{\rm coll}$}}
\newcommand{\npart}    {\mbox{$N_{\rm part}$}}
\newcommand{\nhard}    {\mbox{$\langle n_{\rm h} \rangle$}}
\title{p--Pb Results from ALICE \\
with an Emphasis on Centrality Determination}

\author{Andreas Morsch (for the ALICE Collaboration)}

\address{CERN PH Division, 1211 Geneva, Switzerland}

\ead{andreas.morsch@cern.ch}

\begin{abstract}
New ALICE results concerning particle production at low and intermediate \pT\ in p--Pb collisions at 
$\cmsNN = 5.02\, \tev$ are briefly discussed. Emphasis is given to the determination of centrality in p--Pb 
and their implications for binary scaling of hard processes.
\end{abstract}

The analysis of p--Pb collisions is essential for the study of initial and final state effects
in cold nuclear matter and has the main goal to establish a baseline for the interpretation of the
heavy-ion results \cite{Accardi:2004be}.
In September 2012, an LHC pilot p--Pb run took place
at $\cmsNN = 5.02\, \tev$ followed by a long run in February 2013 delivering 
$\approx 30 \, {\rm nb}^{-1}$ of p--Pb collisions for each experiment - precious
reference data for the Pb--Pb studies, 
but also, as soon turned out, good for a few surprises. 
Several measurements of particle production in the low and 
intermediate $\pT$ region clearly show that p--Pb collisions can
not be explained by an incoherent superposition of p-p collisions and indicate the
presence of collective effects. 
In the first part of this paper,
we will briefly discuss recent ALICE results on this topic. 
The above mentioned studies have been performed as a function of particle multiplicity without 
making any use of the collision geometry e.g. the number of binary collisions \ncoll\ or the number of 
participants $\npart = \ncoll + 1$. 
In order to study nuclear modifications of particle production at high
\pT\ we have to compare particle yields measured in p--Pb to the pp reference spectrum scaled by \ncoll . 
The determination of \ncoll\ and biases on the binary scaling will be discussed in sections 2 and 3, respectively. 
\begin{figure}[htbp]
\begin{minipage}{0.5\linewidth}
\centering
\includegraphics[scale=0.4]{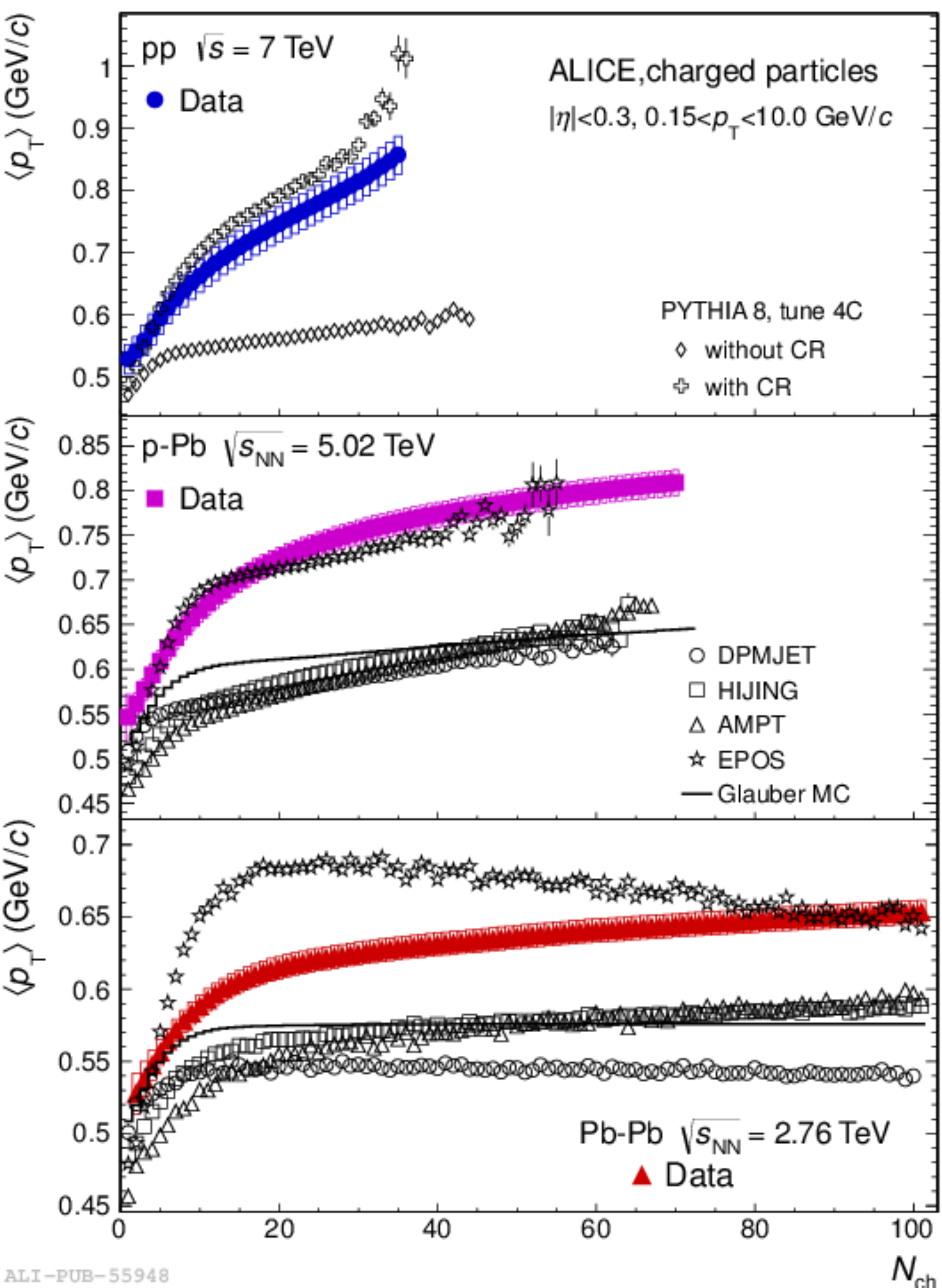}
\caption{$\langle \pT \rangle$ as a function of charged-particle
multiplicity $N_{\rm ch}$ measured in pp (upper panel), p--Pb (middle),
and Pb--Pb (lower) collisions in comparison to model calculations
\cite{Abelev:2013bla}.
%For pp collisions, calculations with PYTHIA~8 with tune 4C are shown with 
%and without the color reconnection (CR) mechanism.
%The p--Pb and Pb--Pb data are compared to calculations with the DPMJET, HIJING,
%AMPT, and EPOS Monte Carlo event generators.
%The lines show calculations in a Glauber Monte Carlo approach.
}
\label{fig:meanpt}
\end{minipage}
\hspace{0.02\linewidth}
\begin{minipage}{0.48\linewidth}
\centering
\includegraphics[scale=0.4]{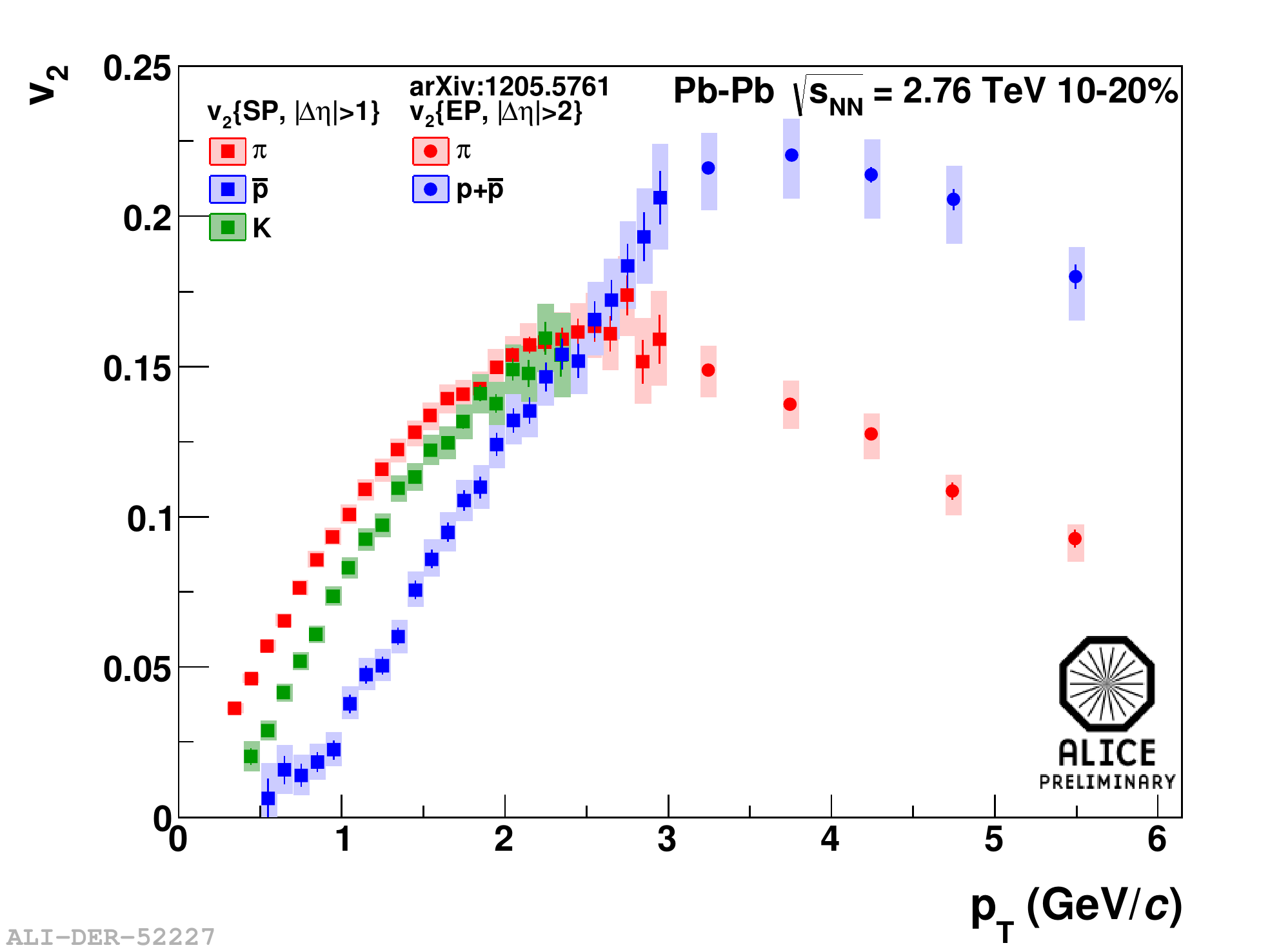}
\includegraphics[scale=0.37]{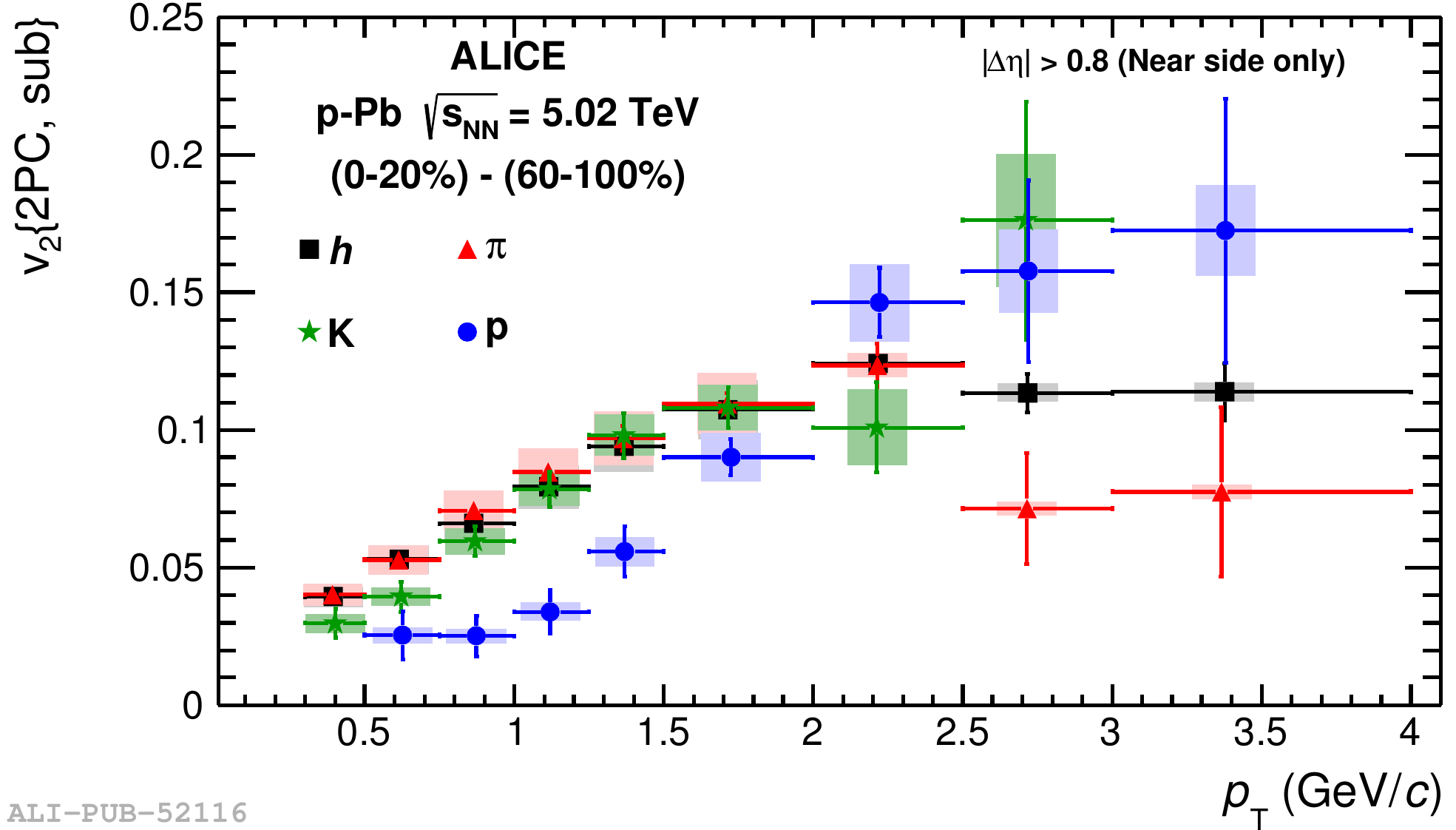}
\caption{For hadrons, pions, kaons and protons:
(upper panel) $v_2$(\pT ) for mid-central Pb--Pb collisions and (lower panel) $v_2\{{\rm 2PC,sub}\}$(\pT ) from 2-particle 
correlations in the 0--20\% multiplicity class after subtraction of the
correlation from the 60--100\% multiplicity class \cite{ABELEV:2013wsa}. 
%Error bars show statistical uncertainties while shadedareas denote systematic uncertainties.
}
\label{fig:pidv2}
\end{minipage}
\end{figure}

\section{Particle production at low and intermediate transverse momentum}
ALICE has measured the average transverse momentum, $\langle \pT \rangle$, as a
function of the charged particle multiplicity, $N_{\rm ch}$, in pp, p--Pb and Pb--Pb collisions 
at $\cmsNN= $ 7~TeV, 5.02~TeV and 2.76~TeV, respectively (Fig. \ref{fig:meanpt}) \cite{Abelev:2013bla}. 
From measurements 
of $\langle \pT \rangle$ in pp at several energies we expect that the collision energy dependence is
weak and, hence, we assume the results for the three collision systems can be directly compared. 
With respect to Pb--Pb, in p--Pb, \meanpt\ shows a much stronger 
increase with multiplicity following the pp data up to $N_{\rm ch} = 14$. 
Note that multiplicities around 14
correspond to typical p--Pb collision, whereas  
pp collisions at this multiplicity are already strongly biased ($N_{\rm ch} > 14$
corresponds to 50\% (10\%) of the p--Pb (pp) cross-section).

In pp, the strong rise of \meanpt\ with multiplicity can be attributed to the presence of color reconnections (CR)
between strings resulting from multiple parton scatterings \cite{Skands:2007zg} which can be interpreted as a 
collective final state effect. 
In other words, the data can not be described by 
an independent superposition of parton scatterings. 
In a similar way, attempts to describe the rise of \meanpt\ in p--Pb
collisions by a superposition of parton scatterings from an incoherent superposition of pp collisions 
fail suggesting 
that also in this case coherent final state effects are at work. Indeed, the EPOS generator %\cite{Pierog:2013ria}
which includes such effects
can reproduce the p-Pb data, however, it fails to describe peripheral Pb--Pb collisions.
ALICE has also measured the multiplicity dependence of \meanpt\ for identified particles ($\pi$, K, p).
Here, a clear mass ordering $\meanpt_p > \meanpt_K > \meanpt_\pi$ is observed \cite{Abelev:2013haa}, which
is an indication for collective expansion with a common velocity field. 

Further evidence for collective effects in p--Pb results from the study of triggered 2-particle angular
correlations in the azimuthal ($\Delta \varphi$) and pseudo-rapidity ($\Delta \eta$) differences. Already
 analyzing  high-multiplicity p--Pb collisions from the pilot run CMS has reported
the presence of a near-side ridge structure elongated in $\Delta \eta$ \cite{CMS:2012qk}.
Using low-multiplicity events as a reference, ALICE and ATLAS found that the near-side 
ridge actually has an almost perfectly symmetrical
counter-part, back-to-back in azimuth \cite{Abelev:2012cya, Aad:2012gla}
very similar to the momentum anisotropy observed in Pb--Pb, 
where the effect is attributed to collectivity (flow). 
In Pb--Pb, this interpretation is further corroborated by the mass dependence of the elliptic flow
coefficient $v_2$. In mid-central Pb--Pb collision (10-20\%) and $\pT < 2\, \gev$, a clear
mass ordering $v_2^\pi > v_2^K > v_2^p$ is observed. Also the slopes are different and this leads to a
crossing-point at around $2.5 \, \gev$. The same behavior is observed in p--Pb
collisions (Fig. \ref{fig:pidv2})  \cite{ABELEV:2013wsa}.
\section{Centrality in p-Pb}
 ALICE has measured the nuclear modification factor \RpA\ for minimum bias p-Pb collisions and 
 it is found to be unity for \pT\ above $\approx 6 \, \gev$ \cite{ALICE:2012mj}. 
 For minimum bias collisions, \ncoll\ is fixed 
 by the total p--Pb and proton--nucleon (p--N) cross-sections: 
 $N_{\rm coll}^{\rm MB} = 208 \, \sigma_{\rm pN}/\sigma_{\rm pPb} = 6.9$. 
 How can \ncoll\ be determined for different centrality classes? In general, centrality
 is defined via centrality estimators that depend monotonically on the number of collisions, e.g.
 multiplicity and summed energy in a certain pseudo-rapidity range. In contrast to Pb--Pb collisions, for p--Pb
 the multiplicity fluctuations for a fixed \ncoll\ are large with respect to the relatively small
 range  of \ncoll\ (typically between 1 and 16). The presence of large fluctuations can bias the p--N collisions 
 themselves. Hence, for each centrality class we have to answer two independent questions: {\it What is the mean number 
 of collisions?} and  
 {\it How much are the p--N collisions biased?} Let us start with the answer to the first question.
\begin{figure}[htp]
\centering
\begin{minipage}{0.48\linewidth}
%\vspace{1.5cm}
\includegraphics[scale=0.4]{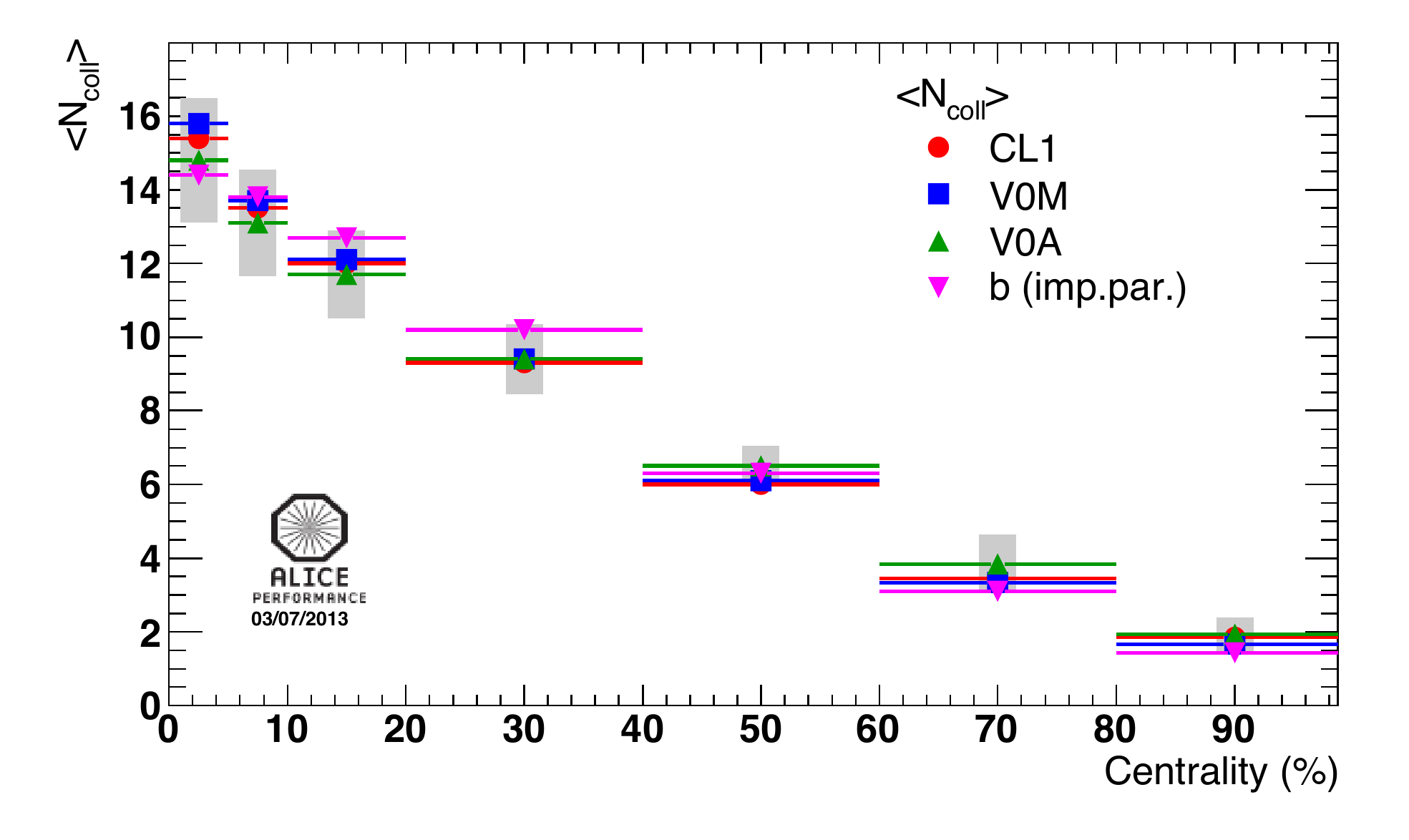}
\caption{Average \ncoll\ as a function of centrality for different estimators. Shaded areas
represent the systematic uncertainties for the V0A estimator.}
\label{fig:ncoll}
\end{minipage}
\hspace{0.2cm}
\begin{minipage}{0.48\linewidth}
\includegraphics[scale=0.4]{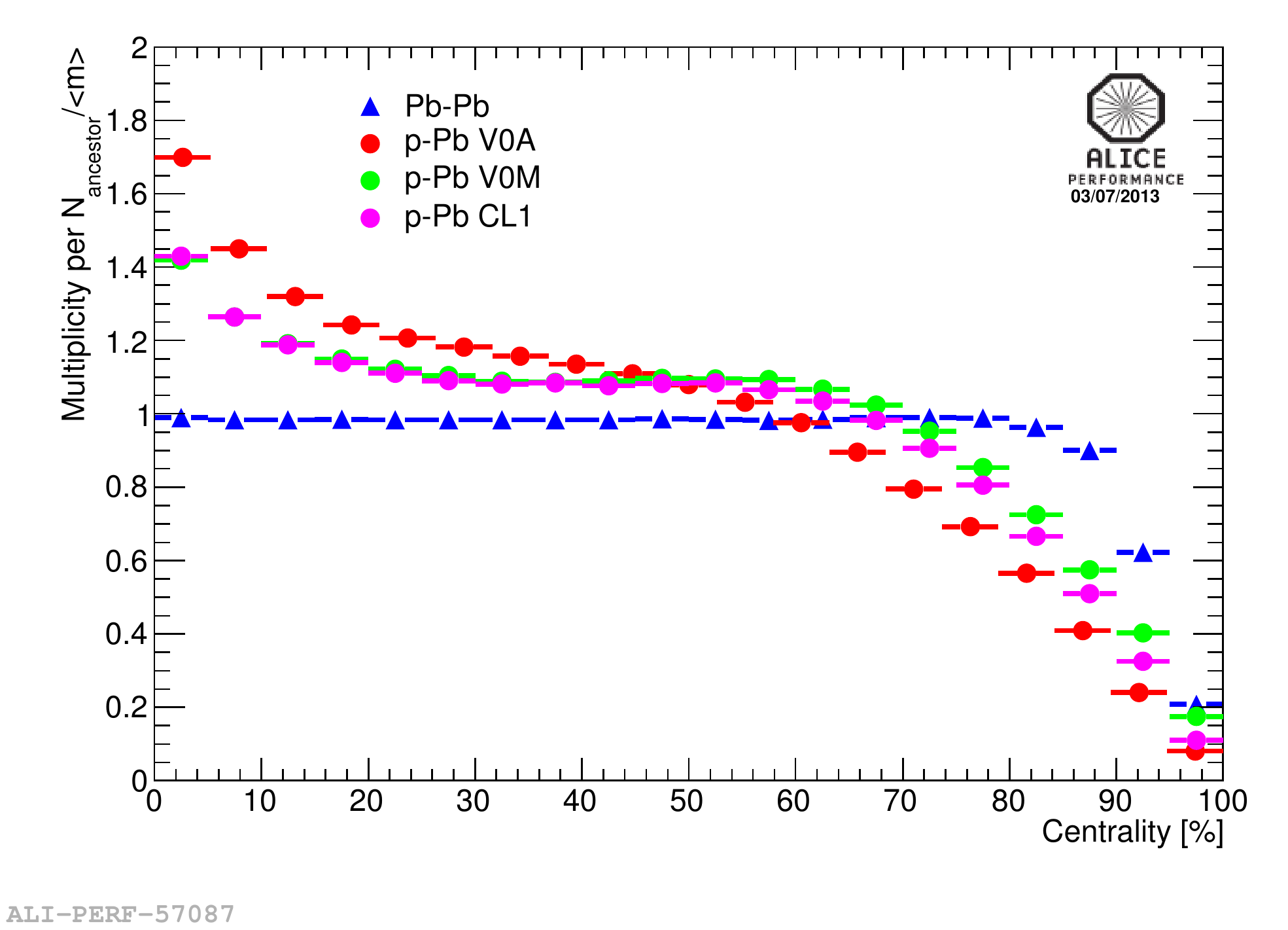}
\caption{Ratio of multiplicity per ancestor to the NBD mean multiplicity.}
\label{fig:bias}
\end{minipage}
\end{figure}
ALICE employs various sub-detector systems covering disjunct pseudo-rapidity ranges to estimate the centrality.
The number of tracks are reconstructed and counted using ITS and TPC  covering $|\eta| < 0.9$. Clusters are counted in the
innermost layers of the ITS, the Silicon Pixel Detector, in $|\eta| < 2.0$ and $|\eta| < 1.4$.
A pair of scintillator hodoscopes, VZEROA and VZEROC, measure charged particle multiplicity in  
the ranges $2.8 < \eta < 5.1$ and $-3.7 < \eta < -1.7$, respectively.
The zero degree calorimeters (ZDC) measure forward proton and neutron energies at beam rapidity. 
The centrality estimators discussed in this paper are: 
{\it CL1}: number of clusters in the 2nd pixel layer, {\it V0A}:
the VZEROA multiplicity, {\it V0M}: the sum of VZEROA and VZEROC multiplicities, {\it ZNA}: the ZDC neutron 
calorimeter energy in direction of the Pb-beam.   

Using these estimators we are sensitive to the reaction products of p-N collisions, the Pb fragmentation products that go
mainly in the direction of the Pb beam ({\it V0A}) and the so called slow nucleons from evaporation and knock-out that are emitted 
into the very
forward directions and are detected by the zero degree calorimeters.
For centrality estimation, particle production in the central part is modeled by negative binomial distributions (NBD) and 
forward slow nucleon production by fragmentation models for evaporation and knock-out \cite{Sikler:2003ef}.

As for Pb--Pb collisions, in p--Pb, we define centrality classes as percentiles of the multiplicity distributions.
In order to extract for each class $\langle \ncoll \rangle$, we use the Glauber fit approach with \npart\ as the number of 
particle sources (ancestors) \cite{Abelev:2013qoq}.
%\npart\ is first obtained from a Glauber Monte Carlo and is equal to the number of particle sources, the ancestors. 
%For each ancestor the multiplicity is obtained from a negative binomial distribution. 
%The procedure is iterated varying the NBD parameters until the best fit to the measured multiplicity distribution has been
%obtained.
Fig. \ref{fig:ncoll} shows $\langle \ncoll\ \rangle $ as a function of centrality for the different estimators.
The variation of $\langle \ncoll \rangle$ between different estimators is small and of similar magnitude as the systematic 
error obtained by varying the Glauber parameters and from a closure test using HIJING \cite{Gyulassy:1994ew}
simulations.
\section{Bias on binary scaling}
Let us now tackle the second question: {\it How biased are {\rm p--N} collisions for a given centrality class?}
Compared to Pb--Pb collisions, we find that in p--Pb collisions the correlation between the centrality estimator 
and \ncoll\ is very loose. In other words the same value of \ncoll\ can contribute to several adjacent centrality classes.
So, what exactly distinguishes two centrality classes for the same \ncoll ? Is this relevant for other physics observables?
The Glauber Monte Carlo itself can give us an indication on the strength of these biases. 
Fig. \ref{fig:bias} shows the ratio of the generated multiplicity per ancestor to the mean multiplicity of the NBD. This 
ratio is
constant in case all collisions are unbiased. What one sees is that in p--Pb the ratio is above unity for central collisions 
and below unity for peripheral collisions, whereas in Pb-Pb the bias is much smaller and restricted to very peripheral 
collisions.
Also the mean p--N impact parameter, $b_{\rm NN}$, can be extracted from the Glauber MC. For peripheral collisions, it is 
larger than the average. Note that in some models large $b_{\rm NN}$ correspond to softer than average interactions.

Multiplicity fluctuations described by the NBD have no immediate dynamical interpretation. However, models based on multi-parton
interaction 
include intrinsically a fluctuating number of particles sources, the hard scatterings. 
For example, in HIJING \cite{Gyulassy:1994ew}, the mean number of hard scatterings, \nhard ,
for a p--N collision is obtained from an impact parameter dependent p--N overlap function, $T_{\rm N}$, and the hard
cross-section, $\sigma_{\rm hard}$, via
$\nhard = T_{\rm N} (b_{\rm NN}) \sigma_{\rm hard}$. 
The number of scatterings, $i$, itself follows a Poissonian distribution 
$p_{i} = {\nhard ^i} / {i!} \exp{(-\nhard ) \rangle}$.
Hence, in these models, there is a natural link between multiplicity fluctuations and the number of hard scatterings.
 \begin{figure}[htp]
\centering
\begin{minipage}{0.48\linewidth}
\includegraphics[scale=0.4]{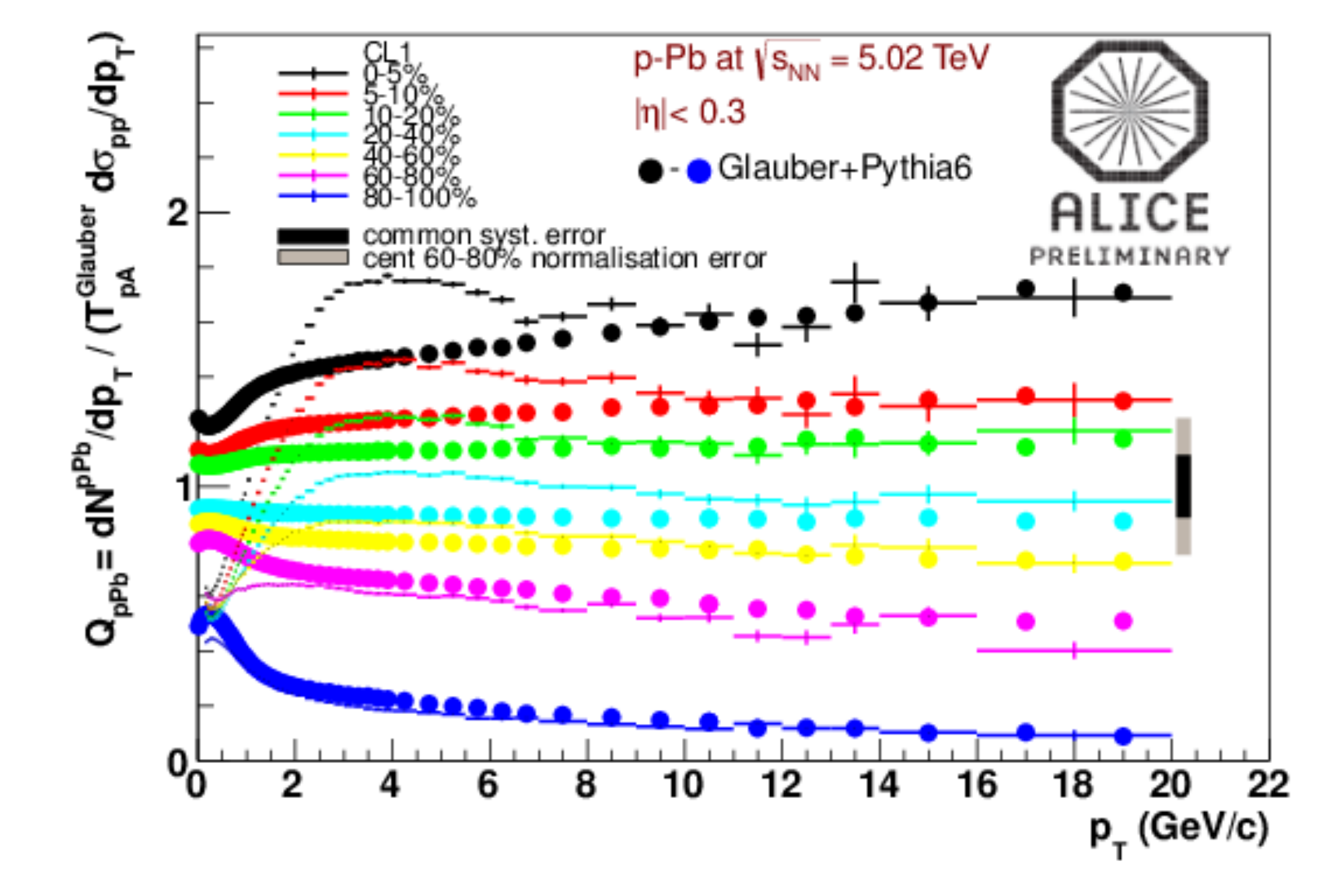}
\end{minipage}
\begin{minipage}{0.48\linewidth}
\includegraphics[scale=0.4]{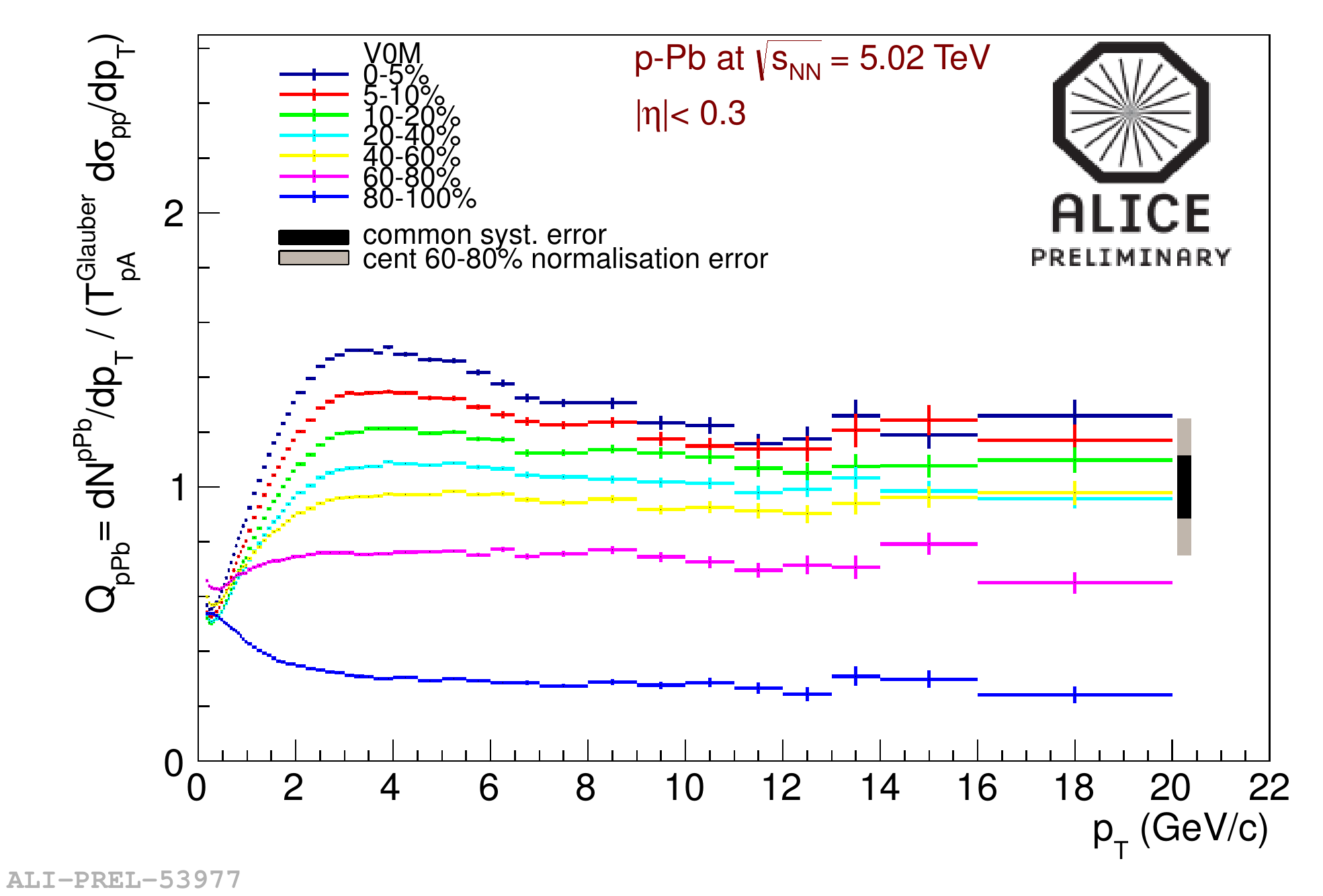}
\end{minipage}
\begin{minipage}{0.48\linewidth}
\includegraphics[scale=0.4]{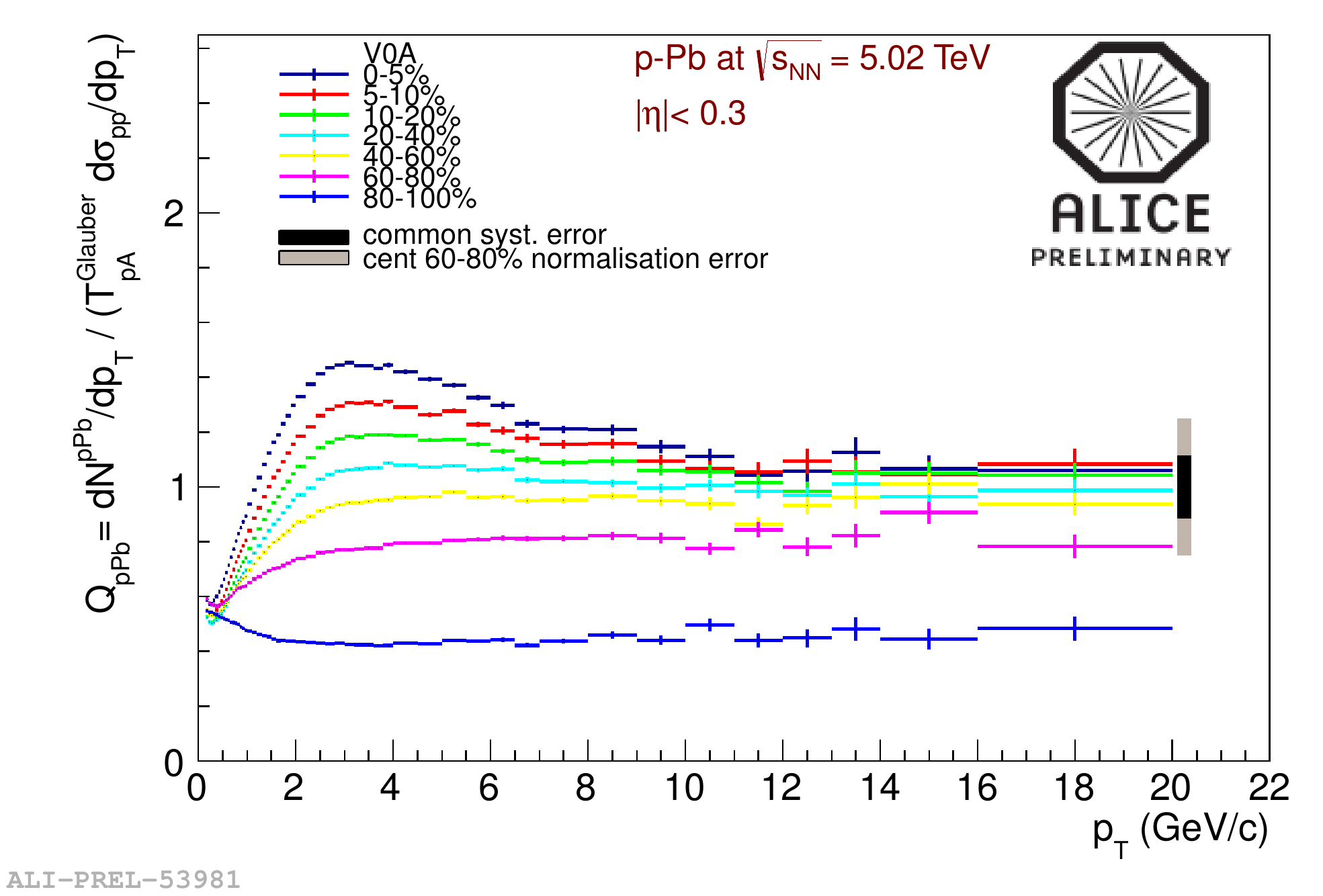}
\end{minipage}
\begin{minipage}{0.48\linewidth}
\end{minipage}
\caption{\QpA (\pT ) for three centrality estimators CL1, V0M and V0A. The CL1 results are compared to a Pythia6 simulation (solid circles, see text)}
\label{fig:qpa}
\end{figure} 
\\ \\
The number of binary p--N collisions is used to scale the reference pp yields and obtain the nuclear
modification factor. However, from the discussion above we have at least two new elements:
For a given centrality class, hard processes scale with $\langle \ncoll \rangle \nhard_{pN} / \nhard_{pp}$.
Further, for a given p--Pb impact parameter, $b$, \nhard\ depends on the average p--N impact parameter.
This is mainly important for peripheral collisions. Here, the multiplicity estimator acts also
as a veto on hard processes which contribute to the overall multiplicity (jet-veto). 
Being aware of this biases we define the observable 
$\QpA  = \dd N^{\rm pPb}/ \dd \pT / (\langle \ncoll \rangle \dd N^{pp}/\dd\pT) =  \dd N^{\rm pPb}/ \dd \pT / (\langle T_{\rm pPb} \rangle \dd \sigma^{pp}/\dd\pT)$,
which is not equal to \RpA , since we do not take into account the bias on the mean number of hard
scatterings.
Fig. \ref{fig:qpa} shows \QpA (\pT ) for the CL1, V0M and V0A centrality estimators. 
For all centrality classes \QpA\ strongly deviates from unity at high \pT . However the spread between different centrality classes
reduces when moving from CL1 to V0M to V0A, increasing the $\eta$-separation between the estimator and the \pT\ measurement. The smallest bias is expected for the ZNA-estimator, the analysis of which is still in progress.
For the most peripheral collisions with CL1, there is a clear indication for a jet
veto bias: \QpA\ has a significant negative slope due to the fact that the contribution of jets to the overall multiplicity increases 
with \pT . The slope is reduced for V0M and absent for V0A. 
The CL1 results are compared to a Pythia6 \cite{Sjostrand:2006za} simulation, where \ncoll\ pp-collisions have been superimposed 
with a probability proportional  to the Glauber \ncoll\ distribution. 
As for the data, the centrality has been obtained from the charged particle multiplicity in $|\eta| < 1.4$
and $\langle \ncoll \rangle$ is directly obtained from the Monte Carlo. Quite surprisingly this simple model 
can reproduce the large bias at high \pT . It also agrees with the low-\pT\ region of the most peripheral collisions
and reproduces the jet-veto bias. For all other centralities, it does not reproduce the Cronin-like enhancement at 
$\approx 4 \, \gev$ and the low-\pT\ region is overestimated. 
%This expected since this region is governed by \npart -scaling.
The latter is expected since in this region also the minimum bias $\RpA$ is below unity.

{\bf In summary}, in pp and p-Pb a strong increase in \meanpt\ with multiplicity is
observed. In models for pp this is understood as a consequence of color reconnections
between strings produced in multiple parton interactions raising 
the question whether similar effects are at work in p--Pb or whether it is the result of collective flow
or initial state effects.
The $v_2$ coefficients obtained from “jet”-subtracted identified 2-particle correlations show a mass
ordering and crossing of $v_2$ of protons and pions.
The pattern is reminiscent of Pb-Pb collisions, where this is effects is attributed to hydrodynamic flow.
Centrality estimators based on multiplicity measurements in
$|\eta| < 5$ induce a bias on the hardness of the p--N collisions that can
be quantified by the number of hard scatterings per p--N collision.
Low (high) – multiplicity p--Pb corresponds to lower (higher) than average number of
hard scatterings. The comparisons to incoherent superposition of p--N collisions have to be
performed including this bias.
% 1 more line left :-)

\end{document}